# Evaluating the Potential Functions of an International Institution for AI Safety – An Institutional Analysis of the IAEA and IPCC, in the Context of Recent Trends in AI Safety

**Arcangelo Leone de Castris** *(The Alan Turing Institute)*
**Christopher Thomas** *(The Alan Turing Institute)*

## Abstract

Governments, industry, and other actors involved in governing AI technologies around the world agree that, while AI promises to deliver tremendous benefits to society, appropriate guardrails are required to mitigate the risks related to this technology. International institutions – including the OECD, the G7, the G20, UNESCO, and the Council of Europe – have already started developing frameworks for ethical and responsible AI governance. While these are important initial steps, alone they fall short of addressing the need for institutionalised international processes to identify and assess potentially harmful AI capabilities. Contributing to the ongoing conversation on how to address this gap, this chapter reflects on the opportunity to establish an international institution for AI safety. Based on the analysis of existing international institutions addressing safety considerations in adjacent policy areas and the newly established national AI safety institutes in the UK and US, this chapter identifies a list of concrete functions which could be performed by an international institution for AI safety. Understanding the impact of international governance institutions developed to address safety risks in other policy domains provides insights into the potential tools at the disposal on an international AI Safety Institution, be it a wholly new international body or a network of national safety institutions and/or other existing fora. Remaining agnostic to its institutional structure, our analysis suggests that the potential functions for an international institution for AI safety can be categorised into three domains: (a) technical research and cooperation, (b) safeguards and evaluations, and (c) policymaking and governance support.

# 1    Introduction

In response to the rapid commercialisation of generative AI systems, 2023 saw several calls for global AI safety regulation from prominent figures across industry and policy (Chowdhury, 2023; Altman *et al*., 2023; Suleyman & Schmidt, 2023; Milmo [Hassabis,] 2023; Prime Minister's Office, 10 Downing Street, 2023; Gutierres, 2023). As a result, the relevant scholarly conversation has increasingly focused on questions related to the establishment of international institutions for the governance of advanced AI (Ho *et al*., 2023; Marcus & Reuel, 2023; Roberts *et al*., 2023; Veale *et al*., 2023). The run-up to the AI Safety Summit hosted by the UK in November 2023, in particular, saw calls for international institutions for AI safety analogous to the International Panel on Climate Change (IPCC) and the International Atomic Energy Agency (IAEA). However, critics argue that drawing inaccurate parallels between different policy domains risks distorting conversations around AI regulation and that an AI safety institution modelled on the IPCC or IAEA examples would inherit the same problems that have limited the impact of these institutions (Broughel, 2023; Afina & Lewis, 2023; Stewart, 2023; De Pryck *et al*., 2022). However, even if these examples are not directly applicable to the international governance of AI, considering governance models established in adjacent policy domains could still provide valuable lessons for this nascent field (Milmo [Hassabis] 2023).

To investigate this assumption, in this chapter we offer an institutional analysis of the International Atomic Energy Agency (IAEA) and the International Panel on Climate change (IPCC) (Steinmo *et al*., 1992). We investigate the functions and policy outcomes of these institutions to determine their relevance to governance processes for AI safety. We compare these findings with the latest efforts to establish national AI safety institutes in the UK and US. By contextualising the analysis of these case studies within  current developments in AI safety and the broader AI governance landscape, we extend the current literature, offering an interpretation of the potential functions that an international AI safety institution could perform. This analysis is underpinned by a mapping review of the relevant literature, situating AI safety within the broader fields of AI ethics and governance, and providing a clear target for our analysis (Booth, 2016). Limited space in this chapter prevents us from offering a broader comparison to other international institutions (e.g. the International Financial Stability Board) and more recently emerging AI safety institutions (e.g., the newly established AI safety institutes in Canada, Singapore, and South Korea). These remain important avenues for further research.

The first section of this chapter sets out the relationship between contemporary interpretations of AI safety focused on 'advanced AI' and concepts of AI ethics and governance (Jobin *et al*., 2019), situating the establishment of AI safety institutions within the broader global AI ethics and governance landscape. In Sections II and III, respectively, we offer a case study analysis of the International Atomic Energy Agency (IAEA) and the International Panel on Climate Change (IPCC), exploring their successes and failures as well as the relevance of these models for AI safety. In Section IV we assess the potential of these analogies for informing the development of AI safety institutions. The IPCC and the IAEA are the focus of this short chapter due to their prominence in the global conversation around AI safety. Hence, we argue that there is a need for a robust and timely analysis of their potential pros and cons. In section V, we analyse recent attempts by the UK and US to establish AI safety institutions, assessing the functions of these institutions at the national level and comparing these efforts to the insights drawn from the IPCC and IEAE case studies. In section VI, we synthesise our analysis to propose potential functions of an international AI safety institution.

## 2    Situating 'safety' within the context of AI ethics and governance

The conversation around AI safety is closely connected to the two broader concepts of AI ethics and AI governance. AI ethics encompasses:

*"a set of values, principles, and techniques that employ widely accepted standards of right and wrong to guide moral conduct in the development and use of AI technologies" (Leslie, 2019).*

Within AI ethics, "safety" is used to represent one of the principles that can guide moral conduct in the pursuit of ethical AI. The translation of abstract ethical principles into actionable organisational processes is one component of the broader category of AI governance, which refers to:

*"a system of rules, practices, processes, and technological tools that are employed to ensure an organization's use of AI technologies aligns with the organization's strategies, objectives, and values; fulfills legal requirements; and meets principles of ethical AI followed by the organization"* (Mäntymäki *et al.* 2022).

It is important to situate proposals for an international AI safety institute within the AI governance context, for two reasons: first, this context enables us to specify the role of an AI safety institute relative to other, complementary mechanisms for AI governance; secondly, this context is important because AI safety regulation won't happen in a vacuum, rather it will sit within a broader corpus of AI standards, principles, and international agreements, as well as a broader array of regulatory access points beyond AI governance such as platform governance, data protection, procurement, international trade agreements, etc. As such, AI safety can be considered as a subset of AI governance.

For the purposes of this chapter, we characterise AI safety as a combination of both negative and positive conditions: (a) *"the absence of unacceptable risk or harm caused by the use of AI"*, and (b) "protections against the unacceptable risk or harm caused by the use of AI*"* (Habli, 2023). In this sense, the role of an AI safety institute would be to put in place protections, preventing or mitigating risks posed by harmful AI capabilities. The current conversations around 'AI safety' have been shaped by the recent, rapid commercialisation of foundation models and, as a result, focus primarily on 'advanced' AI capabilities (Marcus & Reuel, 2023). Due to their broad potential applications and rapid adoption, foundational models have intensified the governance debate focused on ensuring that these systems benefit humanity, including by promoting sustainable and equitable development and managing their potentially negative global externalities (Ho *et al.*, 2023). Beyond the scholarly literature, the focus on AI safety has gained significant political traction, with AI safety institutions being set up across the world to address the risks from the "most advanced AI systems" and the uncertainty brought about by systems at the "frontier of development" (DSIT AI Safety Institute, 2024).

At the international level, intergovernmental institutions have already been establishing principles and frameworks to govern AI responsibly. These include the OECDs *AI principles* (2019), Expert reports by the Global Partnership on AI (2023),the G7 *Hiroshima process* for Advanced AI (2023), UNESCO's *AI ethics Recommendation* (2022), the Council of Europe's *Convention on AI and Human rights* (2024), and the UN's *Interim Report on Governing AI for Humanity* (2023). In parallel, consensus-based standards for AI safety are being developed by international Standard Development Organisations (SDOs), including the joint ISO-IEC Sub-committee 42 on Artificial Intelligence 5 and the IEEE Artificial Intelligence Standards Committee.[1]

Also relevant are regulations and governance mechanisms established by national and regional sovereign entities, some of which will set global precedents and may produce extraterritorial effects. For example, the EU AI Act (AIA) will affect natural and legal persons developing or deploying AI systems well beyond the EU borders. In the US, although attempts to pass AI laws at the Federal level are unlikely to be successful, significant legislative efforts have been made at the state level. For instance, Colorado recently passed bill SB24-205, which providing protection to consumers interacting with AI systems, and New Jersey passed bill S1588, which imposes bias and audit requirements on providers of automated employment decision tools. Also, the AI Risk Management Framework developed by the US National Institute for Standards in Technology (NIST) is set to play a relevant role internally as a soft law instrument, with talks already underway exploring its interoperability with the EU AIA. Relevant regulatory efforts are ongoing in several other countries too (Roberts *et al.*, 2023). Finally, in addition to AI-specific regulation, the governance of this technology will also rely on a range complementary regulatory access points, including data protection regulation (ICO, 2023), digital platform regulation (Veale *et al.*, 2023), competition law (CMA, 2023), the protection of human rights (Leslie *et al.*, 2021), and ESG frameworks (Thomas *et al.*, 2023).

Within this complex landscape, however, there is a gap: there are no authoritative international institutionalised functions to independently assess the risks and future developments of AI (UN AI Advisory Body, 2023). With the objective of contributing to the conversation around how to fill this gap, in the following section, we explore how the experiences of the IAEA and IPPC can inform approaches to govern transnational AI policy issues from a safety perspective.

## 3    Nuclear governance and the International Atomic Energy

Created in 1957, in the context of post-war nuclear proliferation, the International Agency for Atomic Energy (IAEA) marked the beginning of global nuclear governance. The IAEA is an intergovernmental forum representing 178 countries and serving as the global focal point for scientific and technical cooperation in the nuclear field.

The IAEA's mission stems from its statute (1956) and the Treaty on Non-Proliferation on Nuclear Weapons (NPT, 1970), which designates the IAEA as its verification agency, and encompasses both "promotional" and "regulatory" functions (Fischer, 1997). The promotional function of the IAEA consists of encouraging and assisting with research, development, funding, planning for nuclear projects, supporting the practical application of atomic energy for peaceful use, and where necessary acting as an intermediary in securing or supplying nuclear equipment and materials (Article III, 1, 2; Article XI,). From a regulatory standpoint, the IAEA establishes and administers safeguards – including inspections, examinations and approval mandates – to ensure that fissionable and other materials, services, equipment, facilities and information are not used to further military purposes (Article III, 5; Article XII). The Agency also establishes safety standards and provides for their application (Article III, 6). Its nuclear governance scope is focused on three priority areas: a) safety and security, b) science and technology, and c) safeguards and verification.

---

[1] ISO/IEC JTC1/SC 42, https://www.iso.org/committee/6794475.html

One key criticism of the IAEA relates to its impartiality as a regulatory body promoting nuclear power (Fischer 1997). Roerlich (2022, 3) explains that the IAEA is "caught in a seeming paradox of sharing nuclear materials, technology and knowledge while aiming to deter nuclear weapons programs". Similar claims have also been backed by quantitative work finding that states receiving technical assistance related to the nuclear fuel cycle through the IAEA are more likely to seek nuclear weapons (Brown & Kaplow, 2014). However, other scholars have argued the contrary, citing the notion of 'positive disarmament', emphasising that "technical cooperation – even in the nuclear field – 'served as an instrument to inhibit the technological trajectories of independent nuclear programs" (Krige & Sarkar, 2018). Even Fischer notes that the supposed "paradox" of nuclear regulation rests on a false dichotomy of promotion vs regulation, and a misconception of the nature and significance of IAEA assistance, which has been incomparable to the vast investments in nuclear power by industrialised nations (Fischer, 1997).

Further criticism of the IEAE references its lack of enforcement powers. For example, the IAEA lost its power to inspect North Korea when the country, having Withdrawn from the IAEA in 1994 and from the NPT in 2003, expelled the last IAEA inspectors (Council on Foreign Relations 2024; Roerlich 2022). Similar criticism has been drawn over the pursuit of nuclear weapons programs in Iran, Pakistan and Israel, which fall outside of the IAEA's authority due to not having signed the NPT (Arms Control Association, 2022). Finally, the IAEA has been called into question over whether its role has been used to legitimise the military activities of nuclear states. For example, critics argued that the US and other permanent members of the UN Security Council benefitted from the IAEA not only as an initiative for peace but also as a way to protect their nuclear hegemony (Roerlich, 2022, 6).

## 4       Climate governance and the International Panel on Climate Change

Established in 1998 by the United Nations Environment Programme (UNEP) and the World Meteorological Organization (WMO), the IPCC is an intergovernmental institution bringing together 195 member states and several Observer Organisations. The IPCC's primary function is to provide governments with regular and independent assessments of climate change science that they can use to inform climate policies. Although relevant from a policymaking perspective, the IPCC's climate science assessments are not 'policy-prescriptive' (Vardy *et al*., 2017; Bolin, 2007).

The highest organ of the Panel is its plenary. Here, government representatives meet to decide on strategic matters including the work programme, structure and mandate of the working groups, and the scope of assessment reports. The plenary also approves the annual assessment report and elects the IPCC Bureau, which provides technical and strategic guidance to the Panel and approves the appointment of the scientific experts proposed by governments and Observer Organisations to prepare the IPCC reports. The IPCC's day-to-day work occurs within its three Working Groups (WGs) and in the Task Force on National Greenhouse Gas Inventories. WG1 is dedicated to assessing the fundamental physical science behind climate change; WG2 assesses to the impact of climate change and adaptation capabilities; WG3 evaluates mitigation strategies to address climate risks. The Task Force focuses on developing methodologies for measuring national greenhouse gas emissions.

The IPCC has been criticised for being exposed to political pressures, especially due to governments' involvement in reviewing the scientists' work and the reporting duties of the Panel towards the so-called National Focal Points, i.e. local points of contact between the Panel and member governments (Henderson, 2007). However, if on the one hand, it is true that the process of adopting assessment reports within the IPCC is partially political in nature, on the other hand the intergovernmental nature of the Panel has been key in ensuring the widely recognised authority of its reports and, as a consequence, their impact on climate policies (Vardy *et al.*, 2017). For instance, the IPCC's first report published in 1990 has been instrumental to the creation of the UN Framework Convention on Climate Change.

Another critique directed at the IPCC relates to potential conflicts of interest within the Panel (Schiermeier, 2010). For instance, in 2010, IPCC chairman Rajendra Pachauri was targeted with pressures to resign due to allegations of ties between the research institute he directed in New Delhi and companies with a vested interest in proving the severity of climate change. These claims, dismissed following an independent investigation, contributed to the adoption of the IPCC conflict of interest policy in 2011. In addition to laying out precautionary measures to prevent conflicts of interest between the IPCC leadership and the private sector, the policy also protects scientists from governments' influence.

## 5       Relevance of the IAEA and IPCC models for an International AI Safety Institute

The IAEA and IPCC models represent two of the most notable examples of how to govern global policy issues. Considering the transnational nature and impact of AI technologies, policy observers have looked at these two models as references to inform possible international AI safety processes. For instance, both models deal in their own ways with questions related to risk assessment, management, knowledge sharing, cooperation and foresight related to processes that can pose significant threats to our society. For example, useful parallels can be drawn between climate change and AI regarding the challenge of dealing with the fundamental uncertainty characterising the quality and impact of related risks. At the same time, however, the problems that the IAEA and IPCC address are substantively different from those raised by AI. While adaptation to exogenous processes is a key strategic priority in climate policy, and controlling nuclear proliferation for military purposes is at the core of the IAEA mission, the essence of AI safety is about ensuring that the companies developing these technologies do so in a safe and responsible way. Therefore, if drawing connections between governance models in different policy domains can help us to think creatively about innovative institutional functions for AI safety, it is important to remember that their distinct problems will require bespoke solutions.

Several policy observers and business leaders have highlighted that the IAEA model could provide a starting point for thinking about an organisation offering expert advice and coordination around AI safety (Chowdhury 2023; Ho *et al*., 2023; Marcus & Reuel, 2023). OpenAI also called for "something like an IAEA for superintelligence efforts," which would "inspect systems, require audits, test for compliance with safety standards, place restrictions on degrees of deployment and levels of security, etc." (Altman *et al*., 2023). On June 12, 2023, an IAEA-based model gained further support as proposals for a global AI watchdog were backed by UN Secretary General António Guterres (2023).

In spite of this support, however, criticism of the IAEA as a model for a global AI safety institution has been drawn from (1) the risk of inheriting the shortcomings of the IAEA itself and (2) the limitations in the analogy between nuclear technologies and AI. Criticisms of the IAEA's dual mandate raise similar questions as to whether an international AI safety institute should engage in assistive activities or play solely a watchdog role. Furthermore, recalling the criticism of the power of nuclear states in the IAEA, there is a similar risk of big tech lobbying efforts influencing favourable terms for an AI safety institute, facilitating or legitimising their activities. Further criticism of the nuclear analogy relates to nuclear regulation's reliance on a bottleneck of scarce nuclear materials, which leave unique, measurable signatures. As such, the means for developing nuclear technologies can be effectively regulated and verified through controlled access, inspection and approval processes. For AI, while data and computational power for training advanced models require heavy investment, AI systems are increasingly deployed across all segments of society in a way that is completely the opposite of nuclear technologies (Afina & Lewis, 2023).

Another limitation of the analogy is the fact that it situates AI within an existential risk framing, whereas the modalities and pathways to societal harm from AI are vastly different from nuclear technologies as well as being less clear (Stewart 2023). An existential risk framing fails to capture the range of risks that AI systems can pose to individual autonomy and dignity, to societal prosperity and equality, and the impacts of these technologies on public trust in critical democratic institutions through the increasingly sophisticated spread of disinformation, for example. In addition, the nuclear analogy arguably provides a limiting regulatory lens by focusing on dangerous models, akin to dangerous nuclear technologies or weapons. On the contrary, to effectively regulate the risks of AI technologies, we cannot disconnect the models from the networks that they are a part of (Hendrix [interview with Michael Veale], 2023).

Complementing the debate around the pros and cons of the IAEA model for international AI governance, several high-profile AI experts and business leaders recently advanced a proposal for an International Panel on AI Safety (IPAIS) informed by the IPCC example (Suleyman *et al.*, 2023). Behind this proposal is the assumption that, for sensible AI regulation to be adopted, it is necessary to first address policymakers' fundamental lack of understanding of AI. In this sense, an independent and expert-led institute with the task of advising governments on the state of AI capabilities, risks, and impact could provide clarity and foresight. In addition to assessing the scientific evidence around AI capabilities and risks, the IPAIS would also develop globally shared methodologies for reporting on and evaluating AI systems based on the example of the IPCC's

Task Force on National Greenhouse Gas Inventories. Finally, the IPAIS would also complement and help coordinate an international network of institutes and bodies researching AI safety, thus contributing to building a community of AI experts and fostering the emergence of new research and other types of collaboration.

Reflecting on the proposed institutional set-up of the IPAIS, some have noted that, although desirable, such an institute would pose the risk of becoming another forum used by the companies driving AI research to influence policymaking (Liu, 2023). Differently from climate change, AI technologies are not a negative externality of human activity studied by independent scientists across the world, but a commercial product developed by the same actors capable of carrying out the most advanced research in the field. As such, thinking about how an IPAIS would look requires also thinking about how to ensure that inputs from the private sector can feed into the assessment and evaluation activities of the Institute, while also preventing powerful AI organisations from exercising undue influence on its working agenda.

It is evident that both the IPCC and IAEA models provide important insights which could inform the institutional setup and functions of an international organisation for AI safety. However, their limited capacity for impact, their perceived susceptibility to political influence, and the limitations of both models to capture the complex sociotechnical nature of AI systems must be kept in mind when thinking about their applicability to the design of an international institution for AI safety. In the following sections, we apply these insights to the analysis of current efforts to establish institutionalised processes for AI safety in the UK and US, highlighting to what extent current efforts draw from existing models, their distinctive features, and how these provide for an indication of what functions an international institute for AI safety could perform.

## 6       Ongoing efforts to institutionalise AI safety

The AI Safety Summit hosted by the UK in November 2023 was arguably the first explicit international effort towards institutionalising AI safety functions. One of the objectives of the Summit was, in fact, to establish a process for international collaboration to address the risks of 'frontier AI.' Although the Summit failed to deliver actionable harmonised processes for addressing risks connected to advanced AI technologies, a noteworthy output of the first AI Safety Summit was the Bletchley Declaration. Signed by 28 countries and the EU, the statement formally recognised the need for a global approach to understanding the impact of AI on society. Among other things, signatories committed to supporting an international network

*"…of scientific research on frontier AI safety that encompasses and complements existing and new multilateral, plurilateral and bilateral collaboration…to facilitate the provision of the best science available for policy making and the public good" (DSIT, 2023)*

Considering that a follow-up AI Safety Summit has been convened in South Korea in May 2024, with France next to take on hosting duties in Autumn 2024, this institutionalised process holds the potential to evolve into the first regular mechanism of intergovernmental cooperation specifically targeted at AI safety. The Seoul Ministerial statement, signed by 27 countries and the EU reaffirmed the need for "collaborative international approaches to respond to rapid advancements in AI technologies and their impact on our societies and economies". 10 countries and the EU also agreed on establishing a network of publicly backed AI Safety Institutes (DSIT, 2024a; DSIT 2024b). However, notably China and Brazil, signatories to the Bletchley declaration, are absent from the Seoul Ministerial Statement, demonstrating the challenges in establishing and maintaining international consensus.

The signatories of the Bletchley Declaration appointed a panel of AI experts to draft a State of AI Science Report which was published in advance of the Korean Summit (Bengio *et al.*, 2024). The report aims to facilitate a shared understanding of the risks connected to frontier AI by assessing and synthesising existing scientific research on the capabilities of these technologies. Like IPCC assessment reports, this study will contribute to informing national and international AI policymaking. In further similarity with the IPCC, the researchers working on the State of AI Science Report are advised by an Expert Advisory Panel made up of the 30 countries who were invited to the AI Safety Summit as well as representatives from the EU and UN. Even if it is still far from representing an institutional equivalent of the IPCC, this initiative could represent the first step toward an IPAIS shaped on the IPCC model.

Meanwhile, more consequential developments are taking shape at the national level: ahead of the Summit, the UK and the US announced the creation of their national AI safety institutes, the UKAISI and USAISI.

The UKAISI has been created to address the need to ensure that the potential risks of advanced AI can be assessed by public interest bodies and has a threefold mandate: (a) to develop a process for and perform AI system evaluations to assess safety-relevant capabilities; (b) to conduct foundational AI safety research to support innovative governance solutions; (c) to facilitate the exchange of information between policymakers, industry organisations, academia and civil society at both national and international levels. Evaluation results, although limited by the current state of scientific progress in the field and bearing no material consequence for the release or sale of the models evaluated, aim to provide insights to inform better decision-making by both public policymakers and businesses. The UKAISI has published the assessment methodology which it has been using to evaluate models.[2] However, the Institute has also come under scrutiny for the efficacy its model evaluations (Henshall, 2024; Narayanan & Kapoor, 2024), and for its lack of access to the latest models (Manancourt *et al.*, 2024).

In the United States, the Department of Commerce established the USAISI under the National Institute of Standards and Technology (NIST). The institute's mandate largely matches with its British counterpart. It is tasked with (a) developing tools and guidelines for assessing potentially dangerous capabilities of AI models and evaluating them accordingly to identify and mitigate risks. By so doing, the USAISI will contribute directly to operationalising NIST's Risk Management Framework. (b) The Institute will also facilitate information sharing and coordination with national and international counterparts. (c) Finally, and somewhat differently from the UK AISI, it will develop technical guidance to support the development and enforcement of regulation. More recently, in February 2024, the U.S. Secretary of Commerce announced that the USAISI will be supported by the AI Safety Institute Consortium. The Consortium is a hub bringing together more than 200 AI companies, public sector bodies, researchers, and civil society organisations to establish a data and knowledge-sharing space for AI stakeholders and provide inputs into the research and evaluation work of the USAISI.

---

[2] UKAISI open-source framework for large language model evaluations
https://ukgovernmentbeis.github.io/inspect_ai/

From a functional perspective, therefore, both the UK and the US AI safety institutes appear to be at least partially informed by a combination of the features characterising the IAEA and IPCC models as discussed above. For instance, the two institutes play a key role in facilitating scientific and technical knowledge sharing. They also promote safety by developing methodologies to assess potentially harmful impacts of AI technologies and, like the IAEA, by testing and evaluating them. They support cooperation and collaboration between a diverse array of stakeholders, counting on network effects to unlock new opportunities and advances in science and innovation. Finally, they contribute to the development of evidence-led governance frameworks by publishing independent and authoritative information on the state of play of AI science, including facilitating foresight on future advanced AI capabilities.

Considering the specificity of the topic they address, the UKAIAI and USAISI also have distinctive functions reaching beyond those of IAEA and IPCC models. These include performing foundational research on AI safety issues and liaising directly with private sector organisations developing advanced AI models. Also, differently from the IAEA, the institutes do not have regulatory functions.

## 7      Potential functions of an international AI safety institute

The UK and US national AI safety institutes represent an important step towards identifying and mitigating the potentially harmful impact of AI on society. However, the more advanced and diffused AI technologies become, the more pressing the need for effective and dedicated international coordination on AI safety will be. With a view to establishing an international institution specifically focused on AI safety, the UKAISI and USAISI provide useful references on the types of functions that such a body could perform, albeit reflecting UK and US perspectives on what an AI safety institute can practically do. Building on the examples offered by the UKAISI and USAISI, our analysis of the IPCC and IAEA models, and gaps identified by the UN Interim Report on AI for Humanity (2023), we detail below a list of the core functions that an international AI safety institution could perform. The functions are clustered around three thematic categories:

a) technical research and cooperation; b) safeguards and evaluations; c) policymaking and governance support.

a) Technical research and cooperation

- *Assessing AI safety science and horizon scanning*

  Regularly assess the state of AI safety science globally, identify current and future risks, and assess their potential impact on society and the viability of solutions proposed to address them.

- *Supporting and enabling independent AI safety research*

  Strategically identify priority research areas by curating a regularly updated research plan and providing financial and technical support to carry out foundational research.

- *Facilitating international multistakeholder cooperation and knowledge sharing*

  Provide a hub for coordinated investigative efforts across a network of national and international stakeholders committed to independent AI safety research, and enable the dissemination of technical information – including safety and security incidents – to support better decision-making of both AI users and developers.

- *Technical support and assistance to improve access, equity, and capability with regard to less technically advanced nations including access to compute, training and test data.*

  Support cooperation with less technically advanced nations to ensure global buy-in, and inhibit technological trajectories outside of agreed norms.

- *Identify and institute mechanisms to streamline input from national agencies on topics within their direct purviews*

  Ensure integration with broader national AI policy initiatives and support the adaptation of national approaches to continuous development and commercialisation of advanced AI systems.

- *Coordinating and aligning approaches between national AI safety institutions*

  Enable the practical implementation of testing and evaluation and the adaptation of these functions to distinct contextual risks around the globe.

b) *Safeguards and evaluations*

- *Developing processes and methodologies for AI model evaluations*

  Develop an internationally shared set of tools and benchmarks for evaluating potentially dangerous capabilities of AI and mitigating risks. These will have to be applied consistently across an international network of national AI safety institutes and other independent organisations performing AI testing to support global harmonisation of the ways AI risks are assessed.

- *Evaluating AI models*

  Perform testing and evaluation of strategically selected AI models developed across the world. Testing and evaluation include red-teaming, human uplift evaluations, AI agent evaluations, etc.

- *Setting AI safety standards and best practices*

  Leverage the network of national AI safety institutes and other organisations committed to AI safety to set globally recognised safety standards and best practices for developing, deploying, and procuring AI technologies.

- *Establishing powers to inspect models*

  Extend current voluntary evaluation functions of national institutions for the purposes of verifying the appropriateness and compliance of organisations' safeguarding practices against agreed standards.

- *Harmonising international standards and coordinating the development of international standards, safety, and risk management frameworks for advanced AI*

  Ensure the streamlined and effective development and application of norms for AI safety.

- *Publishing evaluation best practices and methodologies*

  Harmonise global AI safety evaluation methodologies and approaches, provide global leadership and support knowledge transfer.

- *c) Policymaking and governance support*

- *Informing policymaking by providing evidence-based scientific assessments*

  Synthesise and assess research developed by national AI safety institutes, and disseminate findings across a network of international and national policymakers to support the development of innovative, evidence-based governance solutions.

- *Offering technical guidance to support the development and implementation of norms and rules*

  Develop technical resources to facilitate regulators' rulemaking and enforcement on issues related to AI safety. Fill gaps in national and international approaches, support best practices, and ensure responsibility and accountability are incentivised at the national level.

- *Ensuring Interoperability and alignment with regional, national, and industry norms*

  Enable policymakers to integrate methodologies, standards and recommendations from the AI Safety Institute.

## 8    Conclusion

This chapter contributes to the ongoing scholarly and policy conversation around safety as a component of international AI governance. It addresses the lack of institutionalised international processes to identify and assess potentially harmful AI capabilities by reflecting on what functions an international AI safety institute could perform. Based on the analysis of both existing international governance models addressing safety considerations in adjacent policy areas and the newly established national AI safety institutes in the UK and US, our analysis identifies a list of concrete functions that could be performed at the international level. While we do not imply that creating a new body is the only way forward, understanding the structure of these bodies from a modular perspective can help us to identify the tools at our disposal. These, we suggest, can be categorised under three functional domains: a) technical research and cooperation, b) safeguards and evaluations, c) policymaking and governance support.

This analysis could be extended by future research in at least three complementary directions. First, by considering not only the functional aspect of institutions like the IPCC and IAEA, but also learnings from their institutional design, including questions of funding, membership, stakeholder relationships, and governance. Second, further research should study a wider range of institutions, to gather more diverse perspectives to inform international AI safety regulation. With respect to policy areas adjacent to AI safety, other organisations responsible for coordinating international efforts to mitigate different types of risk include the Financial Action Taskforce (FATF) and the European Organization for Nuclear Research (CERN). And, most importantly, several countries are now setting up national AI safety institutes – including Japan, Singapore and Canada. It will be interesting to analyse how their functions map against those of the UKAISI and USAISI. Finally, it is important to consider that there are several options for establishing institutionalised processes for AI safety at the international level. These include setting up a new international institution for AI safety, consolidating a network of national safety institutions (as is currently being pursued), as well as building on existing fora such as GPAI, the UN or the OECD. The pros and cons of each option require further consideration, and future developments should be closely monitored.

## Acknowledgments

This research has not been supported by any direct or indirect funding sources.

## References

Afina, Y., & P. Lewis (2023), The nuclear governance model won't work for AI. *Chatham House – International Affairs Think Tank*, https://www.chathamhouse.org/2023/06/nuclear-governance-model-wont-work-ai

Agrawala, S. (1998), Context and Early Origins of the Intergovernmental Panel on Climate Change, *Climatic Change*, vol. 39, 605-620.

Agrawala, S. (1998), Structural and Process History of the Intergovernmental Panel on Climate Change, *Climatic Change*, vol. 39, 621-642.

AI Safety Institute, DSIT (2024), *Introducing the AI Safety Institute*, GOV.UK. https://www.gov.uk/government/publications/ai-safety-institute-overview/introducing-the-ai-safety-institute

Altman, S., G. Brockman & I. Sutskever (2023), Governance of superintelligence. https://openai.com/blog/governance-of-superintelligence

Arms Control Association (2022), Timeline of the Nuclear Nonproliferation Treaty (NPT), https://www.armscontrol.org/factsheets/NPT-Timeline

Bengio, Y., *et al.* (2024), *International Scientific Report on the Safety of Advanced AI: Interim Report*, DSIT 2024/009, https://assets.publishing.service.gov.uk/media/66474eab4f29e1d07fadca3d/international_scientific_report_on_the_safety_of_advanced_ai_interim_report.pdf

Bolin, B. (2007), *A History of the Science and Politics of Climate Change: The Role of the Intergovernmental Panel on Climate Change*. Cambridge: Cambridge University Press.

Booth, A. (2016), *EVIDENT Guidance for Reviewing the Evidence: a compendium of methodological literature and websites,* Working paper, DOI:10.13140/RG.2.1.1562.9842

Bommasani, R., D.A Hudson, E. Adeli, *et al.* (2022), On the Opportunities and Risks of Foundation Models, *arXiv,* https://doi.org/10.48550/arXiv.2108.07258

Broughel, J. (2023), Creating An IPCC For AI Would Be A Historic Mistake, *Forbes*. https://www.forbes.com/sites/jamesbroughel/2023/11/10/creating-an-ipcc-for-ai-would-be-a-historic-mistake/

Brown, R., & J. Kaplow (2014), Talking Peace, Making Weapons: IAEA Technical Cooperation and Nuclear Proliferation, *Journal of Conflict Resolution*, 58(3), 402–428, https://doi.org/10.1177/0022002713509052

Chowdhury, R. (2023), AI Desperately Needs Global Oversight. *Wired*, https://www.wired.com/story/ai-desperately-needs-global-oversight/.

Council on Foreign Relations (2024), Timeline: North Korean Nuclear Negotiations. *Council on Foreign Relations*. https://www.cfr.org/timeline/north-korean-nuclear-negotiations.

Council of Europe (2023). *Revised "Zero Draft" [Framework] Convention on Artificial Intelligence, Human Rights, Democracy and the Rule of Law public,* https://www.coe.int/en/web/artificial-intelligence/cai

De Pryck, K., & M. Hulme (Eds.) (2022), *A Critical Assessment of the Intergovernmental Panel on Climate Change.* Cambridge: Cambridge University Press. https://doi.org/10.1017/9781009082099

Department for Science, Innovation, and Technology (2024a), *Seoul Ministerial Statement for Advancing AI Safety, Innovation, and Inclusivity: AI Seoul Summit 2024,* DSIT, https://www.gov.uk/government/publications/seoul-ministerial-statement-for-advancing-ai-safety-innovation-and-inclusivity-ai-seoul-summit-2024/seoul-ministerial-statement-for-advancing-ai-safety-innovation-and-inclusivity-ai-seoul-summit-2024

Department for Science, Innovation, and Technology (2024b), *Seoul Intent Toward International Cooperation on AI Safety Science, AI Seoul Summit 2024 (Annex),* DSIT, https://www.gov.uk/government/publications/seoul-declaration-for-safe-innovative-and-inclusive-ai-ai-seoul-summit-2024/seoul-statement-of-intent-toward-international-cooperation-on-ai-safety-science-ai-seoul-summit-2024-annex

Fischer, D., & I.E.A. Agency (1997), *History of the International Atomic Energy Agency: The First Forty Years,* International Atomic Energy Agency.

G7 Leaders (2023), *G7 Hiroshima Process on Generative Artificial Intelligence (AI): Towards a G7 Common Understanding on Generative AI*, Organisation for Economic Co-operation and Development, https://www.oecd-ilibrary.org/science-and-technology/g7-hiroshima-process-on-generative-artificial-intelligence-ai_bf3c0c60-en

GPAI (2023), *State-of-the-art Foundation AI Models Should be Accompanied by Detection Mechanisms as a Condition of Public Release*, Report, 2023, Global Partnership on AI, https://gpai.ai/projects/responsible-ai/social-media-governance/Social%20Media%20Governance%20Project%20-%20July%202023.pdf

Guttieres, A. (2023), *Secretary-General Urges Security Council to Ensure Transparency, Accountability, Oversight, in First Debate on Artificial Intelligence | Meetings Coverage and Press Releases*, https://press.un.org/en/2023/sgsm21880.doc.htm

Habli, I. (2023), *On the Meaning of AI Safety*, York: The University of York.

Henderson, D. (2007), Unwarranted Trust: A Critique of the IPCC Process, *Energy & Environment*, 18(7/8), 909–928.

Hendrix, J. (2023), Exploring Global Governance of Artificial Intelligence, *Tech Policy Press*, https://techpolicy.press/exploring-global-governance-of-artificial-intelligence

Henshall, W. (2024), Nobody knows how to safety test AI, *Time*, https://time.com/6958868/artificial-intelligence-safety-evaluations-risks/?utm_source=The+EA+Newsletter&utm_campaign=4d912c4c6d-EMAIL_CAMPAIGN_2024_04_16_12_48&utm_medium=email&utm_term=0_-4d912c4c6d-%5BLIST_EMAIL_ID%5D

Ho, L., J. Barnhart, R. Trager, *et al*. (2023), International Institutions for Advanced AI, *arXiv*, accessed 02/01/2024, https://arxiv.org/pdf/2307.04699.pdf

Information Commissioner's Office (2023), *Guidance on AI and data protection*, ICO, https://ico.org.uk/for-organisations/uk-gdpr-guidance-and-resources/artificial-intelligence/guidance-on-ai-and-data-protection/

Internation Agency for Atomic Energy (1956), *The Statute of the IAEA*. IAEA. https://www.iaea.org/about/statute

Jobin, A., M. Ienca & E. Vayena (2019), The global landscape of AI ethics guidelines, *Nature,* DOI:10.1038/s42256-019-0088-2

Kerry, F. C. (2021), Strengthening International Cooperation on AI, *Brookings*, https://www.brookings.edu/articles/strengthening-international-cooperation-on-ai/

Krige, J., & J. Sarkar (2018), US technological collaboration for nonproliferation: Key evidence from the Cold War, *The Nonproliferation Review*, *25*(3–4), 249–262, https://doi.org/10.1080/10736700.2018.1510465

Leslie, D., C. Burr, M. Aitken, *et al.* (2021), *AI,* human rights, democracy and the rule of law: A primer prepared for the Council of Europe, *The Alan Turing Institute*, https://www.turing.ac.uk/news/publications/ai-human-rights-democracy-and-rule-law-primer-prepared-council-europe

Leslie, D., (2019), Understanding artificial intelligence ethics and safety, *The Alan Turing Institute,* https://www.turing.ac.uk/news/publications/understanding-artificial-intelligence-ethics-and-safety

Liu, L. (2023), Letter: Setting rules for AI must avoid regulatory capture by Big Tech, *Financial Times*, accessed 10/01/2024. https://www.ft.com/content/6a1f796b-1602-4b07-88cd-4aa408cf069a.

Manancourt, V., G. Volpicelli, G. & M. Chatterjee (2024), Rishi Sunak promised to make AI safe. Big Tech's not playing ball, *Politico,* https://www.politico.eu/article/rishi-sunak-ai-testing-tech-ai-safety-institute/

Mäntymäki, M., M. Minkkinen, T. Birkstedt, & M. Viljanen (2022), Defining organizational AI governance, *AI and Ethics*, *2*(4), 603–609, https://doi.org/10.1007/s43681-022-00143-x

Marcus, G., & A. Reuel (2023), The world needs an international agency for artificial intelligence, say two AI experts, *The Economist*, https://www.economist.com/by- invitation/2023/04/18/the-world-needs-an-international-agency-for-artificial- intelligence-say-two-ai-experts.

Milmo, D. (2023), AI risk must be treated as seriously as climate crisis, says Google DeepMind chief, *The Guardian*, https://www.theguardian.com/technology/2023/oct/24/ai-risk-climate-crisis-google-deepmind-chief-demis-hassabis-regulation

Narayanan, A., & S. Kapoor (2024), AI Safety is not a model property, *AI Snake Oil* https://www.aisnakeoil.com/p/ai-safety-is-not-a-model-property

National Institute of Standards and Technology (2023), *U.S. Artificial Intelligence Safety Institute*, NIST, https://www.nist.gov/artificial-intelligence/artificial-intelligence-safety-institute

National Institute of Standards and Technology (2023), *Artificial Intelligence Risk Management Framework (AI RMF)*, NIST, https://nvlpubs.nist.gov/nistpubs/ai/NIST.AI.100-1.pdf

Organisation for Economic Cooperation and Development (2019), *AI-Principles Overview,* OECD, https://oecd.ai/en/principles

Prime Minister's Office, 10 Downing Street (2023), *UK to host first global summit on Artificial Intelligence*, GOV.UK, accessed 25 February 2024, https://www.gov.uk/government/news/uk-to-host-first-global-summit-on-artificial-intelligence

Roberts, H., E. Hine, M. Taddeo & L. Floridi (2023), Global AI governance: Barriers and pathways forward, *SSRN Scholarly Paper 4588040*, https://doi.org/10.2139/ssrn.4588040

Roberts, H., M. Ziosi & C. Osborne (2023), A Comparative Framework for AI Regulatory Policy, *Ceimia*, https://ceimia.org/en/projet/a-comparative-framework-for-ai-regulatory-policy/

Roehrlich, E. (2022), *Inspectors for Peace*, Baltimore: Johns Hopkins University Press.

*S1588* (NJ., 2024), https://www.njleg.state.nj.us/bill-search/2024/S1588.

*SB24-205* (Col., 2024). https://leg.colorado.gov/bills/sb24-205

Schiermeier, Q (2010), IPCC flooded by criticism, *Nature,* vol. 463, 596.

Steinmo, S., K. Thelen & F. Longstreth (eds.) (1992), *Structuring Politics: Historical Institutionalism in Comparative Analysis,* Cambridge: Cambridge University Press.

Stewart, I. (2023), Why the IAEA model may not be best for regulating artificial intelligence, *Bulletin of the Atomic Scientists*, https://thebulletin.org/2023/06/why-the-iaea-model-may-not-be-best-for-regulating-artificial-intelligence/

Suleyman, M & E. Schmidt (2023), Mustafa Suleyman and Eric Schmidt: We need an AI equivalent of the IPCC, *Financial Times*, accessed 10/01/2024, https://www.ft.com/content/d84e91d0-ac74-4946-a21f-5f82eb4f1d2d.

Suleyman, M., M.F. Cuellar, I. Bremmer, *et al.* (2023), Proposal for an International Panel on Artificial Intelligence (AI) Safety (IPAIS): Summary, *Carnegie Endowment for International Peace*, accessed 02/01/2024, https://carnegieendowment.org/2023/10/27/proposal-for-international-panel-on-artificial-intelligence-ai-safety-ipais-summary-pub-90862.

The White House (2023), *FACT SHEET: President Biden Issues Executive Order on Safe, Secure, and Trustworthy Artificial Intelligence*, The White House, https://www.whitehouse.gov/briefing-room/statements-releases/2023/10/30/fact-sheet-president-biden-issues-executive-order-on-safe-secure-and-trustworthy-artificial-intelligence/

Thomas, C., H. Roberts, J. Mökander, *et al.* (2023), The Case for a Broader Approach to AI Assurance: Addressing 'Hidden' Harms in the Development of Artificial Intelligence, *SSRN Scholarly Paper 4660737,* https://doi.org/10.2139/ssrn.4660737

United Nations AI Advisory Body (2023), *Interim Report: Governing AI for Humanity*, United Nations; United Nations, https://www.un.org/en/ai-advisory-body

United Nations Office for Dissarmament Affairs (1970), *Treaty on the Non-Proliferation of Nuclear Weapons (NPT)*, United Nations, https://disarmament.unoda.org/wmd/nuclear/npt/

Vardy, M., M. Oppenheimer, N.K. Dubash, *et al.* (2017), The Intergovernmental Panel on Climate Change: Challenges and Opportunities, *The Annual Review of Environment and Resources*, vol. 42, 55–75.

Veale, M., K. Matus, R. Gorwa (2023), AI and Global Governance: Modalities, Rationales, Tensions, *Annual Review of Law and Social Sciences,* vol 19(1), 255–275.

Weiss, L. (2017), Safeguards and the NPT: Where our current problems began, *Bulletin of the Atomic Scientists*, *73*(5), 328–336, https://doi.org/10.1080/00963402.2017.1362906